\documentclass{article}

\usepackage{authblk}
\usepackage[pdftex]{graphicx}%
\usepackage{multirow}%
\usepackage{amsmath,amssymb,amsfonts}%
\usepackage{amsthm}%
\usepackage{mathrsfs}%
\usepackage[title]{appendix}%
\usepackage{xcolor}%
\usepackage{textcomp}%
\usepackage{manyfoot}%
\usepackage{booktabs}%
\usepackage{algorithm}%
\usepackage{algorithmicx}%
\usepackage{algpseudocode}%
\usepackage{listings}%

\usepackage{geometry}

\newtheorem{theorem}{Theorem}
\newtheorem{lemma}[theorem]{Lemma}

\usepackage{hyperref}

\newcommand*{\myblock}[1]{\paragraph{#1.}}

\usepackage{adjustbox}

\newcommand*{\intervalI}{\ensuremath{\mathcal{I}}} 

\usepackage{caption} 
\usepackage{microtype}

\newcommand{\UnaryOperatorS}[2][]{%
	\ifx&#1&%
	\ensuremath{\mathop{}\mathopen{}#2\mathopen{}}%
	\else%
	\ensuremath{\mathop{}\mathopen{}#2\mathopen{}\left(#1\right)}%
	\fi%
}
\newcommand{\UnaryOperator}[2][]{%
	\ifx&#1&%
	\ensuremath{\mathop{}\mathopen{}#2\mathopen{}}%
	\else%
	\ensuremath{\mathop{}\mathopen{}#2\mathopen{}\left(#1\right)}%
	\fi%
}

\newcommand{\Oh}[1]{\UnaryOperator[#1]{\mathcal{O}}}

\newcommand{\Om}[1]{\UnaryOperator[#1]{\mathup{\Omega}}}

\newcommand{\Ot}[1]{\UnaryOperator[#1]{\mathup{\Theta}}}
\newcommand{\Omm}[1]{\UnaryOperator[#1]{\mathup{\Omega}}}

\DeclareMathAlphabet{\mathup}{OT1}{msb}{m}{n}

\newcommand{\argmax}{\operatorname{argmax}}

\newcommand{\abs}[1]{\ensuremath{\left|#1\right|}}

\usepackage{cleveref}

\usepackage{comment}

\usepackage{url}
\usepackage{booktabs}
\usepackage{multirow}
\usepackage{thm-restate}

\definecolor{teigiColor}{HTML}{5700B5}
\newcommand*{\teigi}[1]{{\color{teigiColor}\emph{#1}}}

\newcommand*{\instancename}[1]{\ensuremath{\mathsf{#1}}} 
\newcommand*{\functionname}[1]{{{\renewcommand{\rmdefault}{ptm}\fontfamily{ppl}\selectfont\textrm{\textup{#1}}}}} 

\newcommand*{\dst}{\instancename{dst}}

\newcommand*{\LCP} {\instancename{LCP}}

\newcommand*{\ISA} {\instancename{ISA}}

\newcommand*{\CDAWG}  {\instancename{CDAWG}}
\newcommand*{\RLBWT}  {\instancename{RLBWT}}
\newcommand*{\ST}  {\instancename{ST}}
\newcommand*{\SA}  {\instancename{SA}}

\newcommand*{\varRoot}  {\instancename{root}}
\newcommand*{\varSink}  {\instancename{sink}}

\newcommand*{\DICT}  {\ensuremath{\mathcal{D}}}
\newcommand*{\Stab}{\instancename{Stab}}

\newcommand*{\fnLCP}{\functionname{lcp}}

\newcommand*{\fnLen}{\functionname{len}}
\newcommand*{\fnDist}{\functionname{dist}}
\newcommand*{\fnSuffixLink}{\functionname{suffixlink}}

\newcommand*{\fnStrAncestor}{\functionname{strAncestor}}

\newcommand*{\fnSuccessor}{\functionname{successor}}
\newcommand*{\fnPredecessor}{\functionname{predecessor}}

\newcommand*{\fnPrevSmaller}{\functionname{PSV}}
\newcommand*{\fnNextSmaller}{\functionname{NSV}}

\title{LZ78 Substring Compression in Compressed Space}

\author[1]{Hiroki Shibata\thanks{\texttt{shibata.hiroki.753@s.kyushu-u.ac.jp}}}
\author[2]{Dominik K\"{o}ppl\thanks{\texttt{dkppl@yamanashi.ac.jp}}}

\affil[1]{Joint Graduate School of Mathematics for Innovation, Kyushu University}

\affil[2]{Department of  Computer Science and Engineering, University of Yamanashi}

\begin{document}


\maketitle

\abstract{
  The Lempel--Ziv 78 (LZ78) factorization is a well-studied technique for data compression.
  It and its derivates are used in compression formats such as \texttt{compress} or \texttt{gif}.
  Although most research focuses on the factorization of plain data, not much research has been conducted on indexing the data for fast LZ78 factorization.
  Here, we study the LZ78 factorization and its derivates in the substring compression model, where we are allowed to index the data and return the factorization of a substring specified at query time.
  In that model, we propose an algorithm that works in compressed space, computing the factorization with a logarithmic slowdown compared to the optimal time complexity.
}




\section{Introduction}

The substring compression problem~\cite{cormode05substringcompression}
is to preprocess a given input text $T$ such that computing a compressed version of a substring of $T[i..j]$ can be performed efficiently.
This problem has been originally stated for the Lempel--Ziv-77 (LZ77) factorization~\cite{storer78macro},
but extensions to the generalized LZ77 factorization{~\cite{keller14generalized}}, 
the Lempel--Ziv 78 (LZ78) factorization~\cite{koppl21nonoverlapping} as well as two of its derivates~\cite{koppl24computing},
the run-length encoded Burrows--Wheeler transform (RLBWT)~\cite{babenko15wavelet},
and the relative LZ factorization{~\cite[Sect.~7.3]{dissKociumakaTomasz}}
have been studied.
Given that $n$ is the length of $T$,
a trivial solution would be to precompute the compressed output of $T[\intervalI]$ for all intervals $\intervalI \subseteq [1..n]$.
This, however, gives us already $\Om{n^2}$ solutions to store.

For an appealing solution, we want to be able to index a large amount of data efficiently within a fraction of space;
two criteria (speed and space) that are likely to be anti-correlated.
However, as far as we are aware, the substring compression problem has not yet been studied with compressed space bounds that can be sublinear for compressible input data.
Our main target is therefore a solution that works in compressed space and can answer a query in time linear in the output size with a polylogarithmic term on the text length.

In this paper, we build upon the line of research on LZ78 factorization algorithms that superimpose the LZ trie on the suffix tree~\cite{nakashima15position,fischer18lz,koppl21nonoverlapping,koppl24computing,koppl24lz78derivates},
which all use the \Om{n} space (in words) to store the suffix tree.
Here, we make the algorithmic idea of the superimposition compatible with the compact directed acyclic word graph (CDAWG)~\cite{blumer85dawg}, trading a tiny time penalty with a large space improvement for compressible texts.
Furthermore, 
we generalize the LZ78 substring compression algorithm to work
with any data structure that can perform a limited set of suffix tree operations.
As a result, we obtain alternative solutions for computing the LZ78 substring compression.
We also apply the proposed method for LZ78 substring compression to LZD~\cite{goto15lzd} and LZMW~\cite{miller85variations} compression, which are derivates of the LZ78 factorization.
\cref{tabSolutions} summarizes both known and newly proposed solutions for the substring compression problem of LZ78, LZD, and LZMW.

\begin{table}
        \centering
	\caption{
    Solutions for computing the LZ78 (or alternatively LZD, or LZMW) substring compression for a string of length $n$ over an alphabet of size $\sigma$, with $z$ LZ78 (LZD, or LZMW) factors, $e$ CDAWG edges, $r$ BWT runs, and having substring complexity $\delta$~\cite{christiansen21optimaltime}.
    Extra space means the additional working space required for processing queries in addition to the index.
    All space complexities are measured in words.
 }
	\label{tabSolutions}
	\begin{tabular}{lccc}
        \toprule
				method & \multicolumn{2}{c}{space} & time \\
        \cmidrule(lr){2-3}
							 & index & extra space & query \\
        \midrule
		naive  & ${\Om{n^2}}$ & ${\Oh{z}}$ & ${\Oh{z}}$ \\
		Köppl~\cite{koppl21nonoverlapping} & ${\Oh{n}}$ & ${\Oh{z}}$ & ${\Oh{z}}$ \\
		\cref{thmFinalResult} & ${\Oh{e}}$ & ${\Oh{z}}$ & ${\Oh{z \lg n}}$ \\
		\cref{thmRLBWT} & ${\Oh{r \lg \frac{n}{r}}}$ & ${\Oh{z}}$ & ${\Oh{z \left( \lg \lg \frac{r}{\lg n} + \lg \frac{n}{r} + \lg z \right)}}$\\
		\cref{thmDelta} & ${\Oh{\delta \lg \frac{n \lg \sigma}{\delta \lg n}}}$ & ${\Oh{z}}$ & ${\Oh{z \lg^{4+\varepsilon} n}}$\\
        \bottomrule
	\end{tabular}
\end{table}

Our contribution fits into the line of research focused on data compression with the CDAWG\@.
Given $e$ and $z$ are the number of edges of the CDAWG and the number of LZ78 factors, respectively,
in that line, 
	\cite{belazzougui17cdawg} proposed a straight-line program (SLP), which can be computed in \Oh{e} time taking \Oh{e} words of space.
	  Given an SLP of size $\Oh{g}$, \cite{bannai13converting} showed how to compute LZ78 from that SLP in \Oh{g + z \lg z} time and space.
	  Combining both solutions, we can compute LZ78 from the CDAWG in \Oh{e + z \lg z} time and space.
	Recently, \cite{arimura23optimally} showed how to compute, among others, the RLBWT and LZ77 in \Oh{e} time and space.

Beyond LZ78, variants such as LZD~\cite{goto15lzd} and LZMW~\cite{miller85variations} have been introduced to improve compression performance.
Both factorizations represent each factor as the concatenation of two previous factors.  
Since the length of each factor can grow exponentially in the best case, the lower bound on the number of factors becomes $\Om{\log n}$.  
This is significantly smaller than the lower bound for standard LZ78 factorization, which is $\Om{\sqrt{n}}$.
However, while the standard LZ78 factorization can be computed in an online manner in $O(n)$ time using $O(z)$ additional words of space where $z$ is the number of LZ78 factors,  
computing these two variants within the same additional space is more difficult.  
\cite{badkobeh17two} showed that the greedy algorithm for these variants takes $\Omega(n^{5/4})$ time in the worst case.  
The fastest known solution for this problem runs in expected $O(n + z' \log^2 n)$ time, where $z'$ is the size of the corresponding parsing~\cite{badkobeh17two}.
This gap in the computational complexity motivates the investigation into the possibility of computing these factorizations in compressed space.

We also note that our research can be viewed as a practical application of compressed suffix trees.  
The majority of the queries needed to compute substring compressions correspond to standard suffix tree operations.  
Indeed, the compressed indexes we employ can be seen as specific instances or variants of compressed suffix trees.  
Our approach shows how to maintain the LZ78, LZD, and LZMW tries within this framework, leveraging the capabilities of these compressed structures.  
This perspective not only unifies different parsing algorithms but also broadens the applicability of compressed suffix tree-based methods.

This article is an extended version of our contribution~\cite{shibata24substring} to the SPIRE'24 conference.
Our extension is as follows.
First, we have adapted our techniques for substring compression to LZD and LZMW.
Second, we formulate an abstract data type for the data structure we need for our algorithm computing the substring compression for any of the addressed compression types (LZ78/LZMW/LZD). 
We subsequently provide examples of implementations other than CDAWGs for this abstract data type.
Finally, we provide additional experimental results.

\section{Preliminaries}\label{secPrelimanaries}
With $\lg$ we denote the logarithm~$\log_2$ to base two.
Our computational model is the word RAM model with machine word size $\Om{\lg n}$ bits for a given input size~$n$.
Accessing a word costs $\Oh{1}$ time.
Unless stated otherwise, we measure space complexity in words rather than bits.

Let $T$ be a text of length $|T| = n$ whose characters are drawn from an integer alphabet $\Sigma= [1..\sigma]$ 
with $\sigma \le n^{\Oh{1}}$.
Given $X,Y,Z \in \Sigma^*$ with $T = XYZ$, 
then $X$, $Y$ and $Z$ are called a \teigi{prefix}, \teigi{substring} and \teigi{suffix} of $T$, respectively.
We call $T[i..]$ the $i$-th suffix of $T$, and denote a substring $T[i] \cdots T[j]$ with $T[i..j]$.
A \teigi{longest common prefix (LCP)} of two strings $X$ and $Y$ is the longest prefix $P$ of $X$ and $Y$ that satisfies $X[1..|P|] = Y[1..|P|] = P$.
We denote $\fnLCP(X, Y)$ as the longest common prefix of $X$ and $Y$.
The longest common prefix between two suffixes of the string is also called \teigi{longest common extension (LCE)}.
A \teigi{parsing dictionary} is a set of strings.
A parsing dictionary~$\mathcal{D}$ is called \teigi{prefix-closed} if it contains, for each string $S \in \mathcal{D}$,
all prefixes of $S$ as well.
A \teigi{factorization} of $T$ of size~$z$ partitions $T$ into $z$~substrings $F_1 \cdots F_z = T$.
Each such substring $F_x$ is called a \teigi{factor} and $x$ its \teigi{index}.

\myblock{LZ78 Factorization}
Stipulating that $F_0$ is the empty string,
a factorization $F_1\cdots F_z = T$ is called the \teigi{LZ78 factorization}~\cite{ziv78lz} of $T$ if and only if, for all $x \in [1..z]$,
the factor~$F_x$ is the longest prefix of $T[|F_1\cdots F_{x-1}|+1..]$ such that $F_x = F_y \cdot c$ for some $y \in [0..x-1]$ and $c \in \Sigma$,
that is, $F_x$ is the longest possible previous factor $F_y$ appended by the following character in the text.
The dictionary for computing $F_x$ is $\DICT_x := \{F_y \cdot c : y \in [0..x-1], c \in \Sigma \}$, which is prefix-closed.
Formally, $F_x$ starts at $\dst_x := |F_1..F_{x-1}|+1$ and 
$y = \argmax \{ \abs{F_{y'}} : F_{y'} = T[\dst_x..\dst_x+|F_{y'}|-1]\}$.
We say that $y$ and $F_y$ are the \teigi{referred index} and the \teigi{referred factor} of the factor~$F_x$, respectively.
The LZ78 factorization of $T = \texttt{babac}$ is 
$F_0, F_1, \ldots, F_4 = \epsilon, \texttt{b}, \texttt{a}, \texttt{ba}, \texttt{c}$.
The referred factor of $F_3 = F_1 \texttt{a}$ is $F_1$; $F_3$'s referred index is $1$.

$\DICT_x$ is often implemented by the \teigi{LZ trie}, which represents each LZ factor as a node; the root represents the factor~$F_0$.
The node representing the factor $F_y$ has a child representing the factor $F_x$ connected with an edge labeled by a character~$c \in \Sigma$ if and only if $F_x = F_y c$.
To see the connection of the LZ trie and $\DICT_x$, we observe that adding any new leaf to the LZ trie storing $\{F_1,\ldots,F_{x-1}\}$ 
gives an element of $\DICT_x$, and vice versa we can obtain any element of $\DICT_x$ by doing so.
A crucial observation is that every path from the LZ trie root downwards visits nodes in increasing LZ factor index order.

\begin{figure}[t]
\centering
\includegraphics[width=0.8\textwidth]{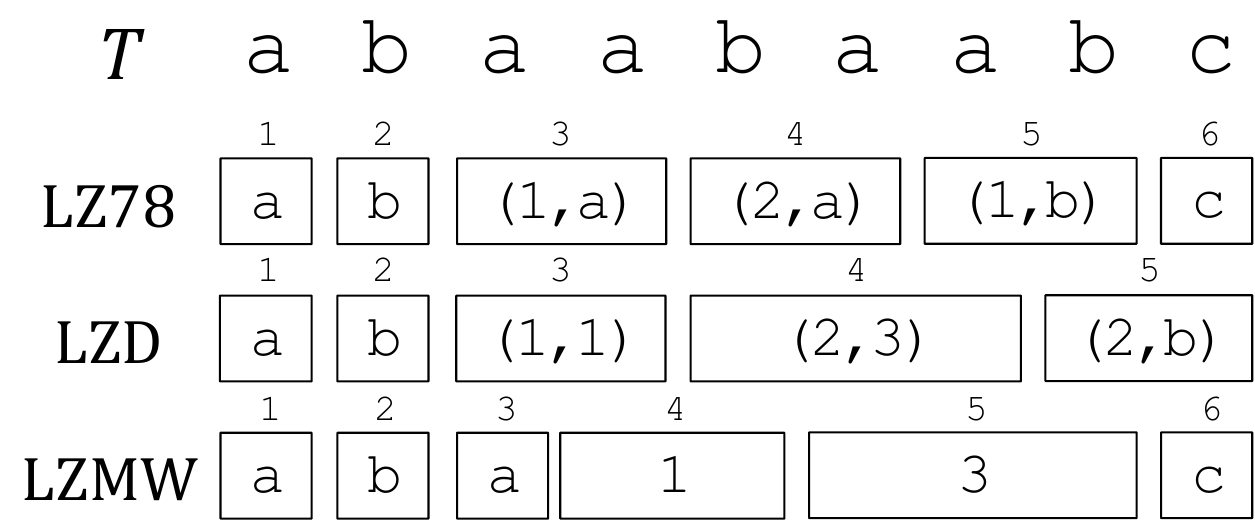} 
\caption{%
    The LZ78, LZD, and LZMW factorizations for $T = \texttt{abaabaabc}$.  
    Each block corresponds to a factor, and the number above each block indicates the factor's index.  
    Blocks labeled with a single character represent that character as a factor.  
    In LZ78, a factor $F_i$ represented by a tuple $(j, c)$ means $F_i = F_j c$.  
    In LZD, a factor $F_i$ represented by a tuple $(j, k)$ means $F_i = F_j F_k$.  
    In LZMW, a factor $F_i$ represented by a single integer $j$ means $F_i = F_j F_{j+1}$.
}
\label{fig_factorizations}
\end{figure}

Lempel-Ziv Double (LZD)~\cite{goto15lzd} and Lempel-Ziv-Miller-Wegman (LZMW)~\cite{miller85variations} factorization are non-prefix-closed variants of LZ78 factorization.
A factorization $F_1 \cdots F_z$ of $T$ is \teigi{LZD} if
$F_x$ is represented by $F_x = G_1 \cdot G_2$ with
$G_1 \in \{F_0, F_1,\ldots,F_{x-1}\}, G_2 \in \{ F_1, \dots, F_{x-1}\} \cup \Sigma$ such that
$G_1$ and $G_2$ are, respectively, the longest possible prefixes of $T[\dst_x..]$ and of $T[\dst_x+|G_1|..]$.
A factorization $F_1 \cdots F_z$ of $T$ is \teigi{LZMW} if $F_x =F_y \cdot F_{y+1}$ with $1 \leq y < x - 1$ or $F_x \in \Sigma$ and all factors are greedily selected by length.
Examples of the LZ78, LZD, and LZMW factorizations are shown in~\cref{fig_factorizations}.
Because non-prefix closed parsings may introduce unary paths in the parse trees,
we need to study LZ tries having unary paths.

\begin{figure}[t]
\centering
\includegraphics[width=0.8\textwidth]{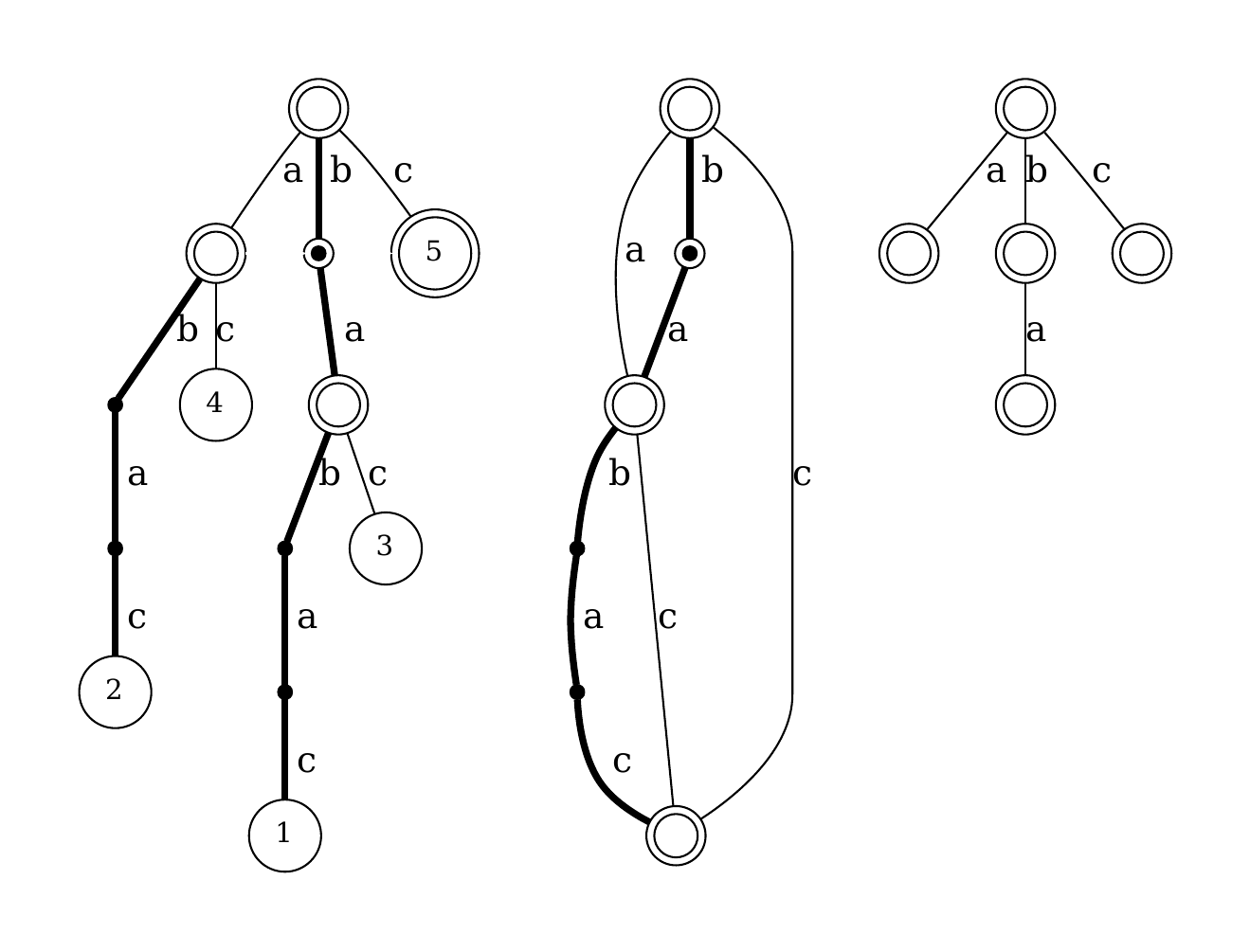} 
    \caption{%
        \ST{} (left), \CDAWG{} (center), and the LZ trie (right) for $T = \texttt{babac}$.
        The LZ78 factorization of $T$ is $F_0, F_1, \ldots, F_4 = \epsilon, \texttt{b}, \texttt{a}, \texttt{ba}, \texttt{c}$. 
        We superimpose the suffix trie and the DAWG on \ST{} and \CDAWG{}, respectively, by drawing implicit nodes with black dots on the edges.
        We additionally encircle vertices corresponding to LZ78 factors (thus showing explicit nodes as double circles).
        Bold and thin lines represent, respectively, the heavy and light edges of the centroid path decomposition.
        The \CDAWG{} sink represents the set of strings $\{ \tt c, ac, bac, abac, babac \}$, which can be read on the \varRoot{}-\varSink{} paths.
        Only a part of these strings are LZ78 factors.
    }
\label{fig_st_lztrie_cdawg}
\end{figure}

\myblock{Suffix Tree}
Given a tree, with an \teigi{$s$-$t$ path} we denote the path from a node $s$ to a node $t$.
All trees in this paper are considered non-empty with a root node, which we denote by \varRoot{}.
The \teigi{suffix trie} of $T$ is the trie of all suffixes of $T$. 
There is a one-to-one relationship between the suffix trie leaves and the suffixes of~$T$.
The \teigi{suffix tree}~\cite{weiner73linear}~\ST{} of $T$ is the tree obtained by compacting the suffix trie of $T$.
The string stored in an \ST{} edge~$g$ is called the \teigi{label} of $g$.
The \teigi{string label} of a node $v$ is defined as the concatenation of all edge labels on the \varRoot{}-$v$ path;
its \teigi{string depth} is the length of its string label.
The leaf corresponding to the $i$-th suffix~$T[i..]$ is labeled with the \teigi{suffix number}~$i \in [1..n]$.
The \teigi{locus} of a substring $S$ of $T$ is the place we end up when reading $S$ from \ST{} starting at \varRoot{}.
The locus of $S$ is either an \ST{} node, or on an \ST{} edge (called an \teigi{implicit node} because it is represented by a suffix trie node).
The left of \cref{fig_st_lztrie_cdawg} gives an example of \ST{}.

Reading the suffix numbers stored in the leaves of \ST{} in leaf-rank order gives the suffix array~\cite{manber93sa}.
We denote the suffix array of $T$ by $\SA$.
The array~$\ISA$ is defined such that $\ISA[ {\SA[i] } ] = i$ for every $i = 1,\ldots,n$.
We also denote the \teigi{LCP array} $\LCP$ of $T$ by $\LCP[i] = |\fnLCP(T[\SA[i]], T[\SA[i + 1]])|$.
Since the \ST{} leaves are sorted in \SA{} order, 
the value $\LCP[i]$ is the string depth of the lowest common ancestor of the $i$-th leaf and $(i+1)$-st leaf in leaf-rank order.
An \ST{} node $v$ can be uniquely represented by an \SA{} range $[i..j]$ such that the $k$-th leaf is in the subtree of $v$ for all $k \in [i..j]$.

\myblock{Centroid Path Decomposition}
The centroid path decomposition~\cite{ferragina08searching} of a tree is defined as follows.
For each internal node, we call its child whose subtree is the largest among all its siblings (ties are broken arbitrarily if there are multiple such children) a \teigi{heavy} node,
while we call all other children \teigi{light} nodes. 
Additionally, we make \varRoot{} a light node.
A \teigi{heavy path} is a maximal-length path from a light node~$u$ to the parent of a leaf containing, except for~$u$, only heavy nodes.
Since heavy paths do not overlap, we can contract all heavy paths to single nodes and thus form the centroid-path decomposed tree
whose nodes are the heavy paths that are connected by the light edges. 
The centroid-path decomposed tree is helpful because the number of light nodes on a path from \varRoot{} to a leaf is \Oh{\lg n}, which means that a path from \varRoot{} to a leaf contains only \Oh{\lg n} nodes.
This can be seen from the fact that the subtree size of a light node is at most half of the subtree size of its parent; thus when visiting a light node during a top-down traversal in the tree, 
at least half the number of nodes we can visit from then on.
Consequently, a \varRoot{}-leaf path in a centroid-path decomposed tree has \Oh{\lg n} nodes.

\myblock{CDAWG}
In what follows, we adapt LZ78-substring-compression techniques to work with the CDAWG instead of \ST{}.
The CDAWG of $T$, denoted by \CDAWG{}, is the minimal compact automaton that recognizes all suffixes of $T$~\cite{blumer85dawg,crochemore97direct}.
The CDAWG of $T$ is the minimization of \ST{}, in which 
 (a) all leaves are merged to a single node, called \emph{\varSink{}}, and
 (b) all nodes, except \varSink{}, are in one-to-one correspondence with the maximal repeats of $T$~\cite{raffinot01maximal},
    where a maximal repeat $S$ is a substring of $T$ having two occurrences $a_1Sb_1$ and $a_2Sb_2$ in $T$ with $a_1 \neq a_2$ and $b_1 \neq b_2$.
When transforming \ST{} into \CDAWG{}, multiple \ST{} nodes can collapse into a single \CDAWG{} node, and we say that such a \CDAWG{} node \teigi{corresponds} to these collapsed \ST{} nodes.
We denote the number of \CDAWG{} edges by $e$.
With $\bar{e}$, we denote the number of edges of the CDAWG of the reverse of $T$.
The number of \CDAWG{} edges $e$ can be regarded as a compression measure.
For highly repetitive text, $e$ can become asymptotically smaller than the text length $n$.
In general, we can bound $e$ with $e \in \Oh{n}$ and $e \in \Om{\lg n}$.
The upper bound is obtained from the fact that the number of \ST{} edges is at most $2n - 1$;  
the lower bound is obtained from the fact that $g \in \Oh{e}$ and $g \in \Om{\lg n}$, where $g$ is the size of smallest grammar that produces $T$~\cite[Lemma 1]{belazzougui17cdawg}.
Furthermore, there is a string family that achieves $e \in \Ot{\lg n}$~\cite{rytter06fibonacci}.
The middle of \cref{fig_st_lztrie_cdawg} gives an example of \CDAWG{}. 

\myblock{RLBWT}
We also adapt the algorithm to work in RLBWT-compressed space.
The \teigi{Burrows--Wheeler transform (BWT)}~\cite{burrows94bwt} of $T$ is the concatenation of the last characters of all rotations of $T$ sorted in lexicographic order.
Since a BWT puts together all the suffixes starting with the same context, it typically has many runs.
Therefore, the \teigi{run-length encoded Burrows--Wheeler transform (RLBWT)}~\cite{babenko15wavelet}, denoted by \RLBWT{}, compresses the text effectively.
It is known that $r \in \Oh{e}$ holds for all strings~\cite{belazzougui15composite}, and a simple string $\texttt{a}^{n-1}\$$ achieves $e \in \Ot{n}$ and $r \in \Oh{1}$.
RLBWT can also be used for compressed indexes.
There is a data structure that efficiently simulates typical suffix tree operations in $\Oh{r \lg \frac{n}{r}}$ space~\cite{gagie20fully}.
Note that the two compression measures $e$ and $r \lg \frac{n}{r}$ are incomparable.
Therefore, an RLBWT-based $\Oh{r \lg \frac{n}{r}}$ space index can be considered as an alternative to a CDAWG-based $\Oh{e}$ space index.

\myblock{Substring Complexity}
The \teigi{substring complexity}~\cite{christiansen21optimaltime} is a repetitiveness measure that can be efficiently computed.
For a given text string $T$ of length $n$, the substring complexity $\delta$ is defined as $\delta = \max \{ d_k(T)/k \mid k \in [1, n] \}$, where $d_k(T)$ is the number of distinct substrings in $T$ of length $k$.
$\delta$ can be computed in linear time and 
is a lower bound for many repetitiveness measures such as the number of edges $e$ of the CDAWG or the number of runs $r$ of the BWT\@.
A string of length $n$ with substring complexity $\delta$ can be represented using $\Oh{\delta \lg \frac{n \lg \sigma}{\delta \lg n}}$ space~\cite{kociumaka23toward}.
Recently, a data structure has been proposed that supports various queries by providing random access on both $\SA$ and $\ISA$ and supporting longest common extension queries, all within the same space complexity~\cite{kempa23collapsing}.

\section{Factorization Algorithm}
The aim of this paper is to compute, after indexing the input text~$T$ in a preprocessing step,
upon request for a provided interval $[i..j] \subseteq [1..n]$,
the LZ78 factorization of $T[i..j]$, 
in compressed space with time bounded linearly in the output size and logarithmic in the text length.
For that, we propose two algorithms, where the first one simulates \ST{} with \CDAWG{}, 
and thus directly applies the techniques of our pre-cursors working on \ST{}.
For the last algorithm, we show how to eliminate the need for the \ST{} functionality to improve the time bounds.

\subsection{Superimposing \CDAWG{}}

In the following, we show how to adapt the \ST{} superimposition by the LZ trie from \cite{nakashima15position} to \CDAWG{}.
The main observation of \cite[Sect.~3]{nakashima15position} is that 
the {LZ} trie is a connected subgraph of the suffix trie containing its root because LZ78 is prefix-closed.
However, the compacted suffix trie, i.e., \ST{}, does not contain all suffix trie nodes.
In fact, the locus of each factor $F_x$ is either an \ST{} node $v$ or lies on an \ST{} edge~$g$.
In the former case, $v$ corresponds to the LZ node $v'$ representing $F_x$ in the sense that both have the same string depth.
In the latter case, the locus of $F_x$ can be witnessed by the lower node the edge~$g$ connects to (by storing information on the length of $F_x$ at that node).
Thus, we can represent the LZ trie with a marking of \ST{} nodes.
The marking is done dynamically while computing the factorization as we mark the locus of each factor after having it processed.
By marking the \ST{} root node, we identify the LZ trie root with the \ST{} root.
To find the factor lengths, we perform a traversal from the leaf $\lambda$ to its lowest marked ancestor,
where $\lambda$ is the leaf whose suffix number corresponds to the starting position of the factor we want to compute.
Thus, we process the leaves in the order of their suffix numbers while computing the factorization.

To translate this technique to \CDAWG{}, we no longer move to different leaves since all leaves are contracted to \varSink{}.
This is no problem if we keep track of the starting position of the factor we want to compute.
However, an obstacle is that a \CDAWG{} node can have multiple parents.
Given we superimpose the LZ trie on \CDAWG{} such that an explicit LZ (trie) node $v$ is stored in its corresponding \CDAWG{} node $v'$.
Unlike the case for \ST{}, we generally have no information on the actual length of the string of $v$ because $v'$ can have multiple paths leading to \varRoot{}.
\cref{fig_st_lztrie_cdawg} presents an example for which we cannot superimpose the LZ trie on \CDAWG{}.
In what follows, we propose two different solutions.

\subsection{First Approach: Plug\&Play Solution} \label{FirstAlg}
Our solutions make the idea of the superimposition more implicit by modeling the LZ trie with a weighted segment tree data structure whose intervals correspond to \ST{} nodes.
In detail, we augment each LZ trie node with an \SA{} range. 
For explicit LZ trie nodes having a corresponding \ST{} node $v$, its \SA{} range is the \SA{} range of $v$. 
Otherwise, its range is the \SA{} range of the \ST{} node directly below.
\SA{} ranges of LZ trie nodes can be nested but do not overlap due to the tree topology of \ST{}.
This makes it feasible to model the lowest marked ancestor data structure used in the precursor algorithms 
with a weighted segment tree data structure that represents each LZ trie by its \SA{} range and its LZ index as weight.
For the example in \cref{fig_st_lztrie_cdawg}, we end by storing the weighted intervals
$(0, [1..5]), (1, [3..4]), (2, [1..2]), (3, [3..4]), (4,[5..5])$, where the first components denote the weights (i.e., the factor indices).
In particular, we can make use of the following data structure for a \teigi{stabbing-max query}, i.e.,
for a given query point $q$, to find the interval with the highest weight containing $q$
in a set of weighted intervals.

\begin{lemma}[{\cite{yang03aggreagte}}] \label{lemStabbingQuery}
    Given a set of $z$ intervals in the range $[1..n]$ with weights in $[1..n]$,
    there exists a linear-space data structure that answers stabbing-max queries and supports the insertion of a weighted interval in $\Oh{\lg z}$ time. 
\end{lemma}

With the data structure of \cref{lemStabbingQuery}, it is already possible to compute the LZ78 factorization without constructing the LZ trie in \Oh{z \lg z} time with \ST{}.
For that we maintain the intervals of all computed LZ factors in an instance \Stab{} of this data structure such that we can identify the factor index by the returned interval.
We additionally index $F_0$ with interval $[1..n]$ and weight $0$ for determining non-referencing factors.
By doing so, given we want to compute a factor $F_x = T[\dst_x..\dst_x+|F_x|-1]$, 
we can determine its reference $y$ by querying \Stab{}.
If $y > 0$, then $F_x = F_y \cdot T[\dst_x+|F_y|]$.
It is left to determine the interval of $F_x$, which we need to add to \Stab{}.
For that, we find the locus of $F_x$ in the suffix tree, which can be done with a weighted level ancestor data structure in constant time~\cite{gawrychowski14weighted,belazzougui21weighted}.

This approach can be directly rewritten for \CDAWG{}.
To this end, we make use of the \Oh{e + \bar{e}}-words representation of \cite{belazzougui15composite} and \cite{belazzougui17cdawg},
which represents an \ST{} node $v$ with \Oh{\lg n} bits of information, namely:
(a) $v$'s corresponding \CDAWG{} node, (b) the string length of $v$, and (c) $v$'s \SA{} range.
Their representation supports the following \ST{} operations:
  (a) \fnSuffixLink($v,i$) returns the \ST{} node after taking $i$ suffix links starting from $v$, in \Oh{\lg n} time;
 (b) \fnStrAncestor($v,d$) returns the highest ancestor of an \ST{} node $v$ with string depth of at least $d$, in \Oh{\lg n} time.

Furthermore, it is known that the number of runs $r$ in the BWT is upper-bounded by $e$~\cite{belazzougui15composite}. 
Hence, in $\Oh{e}$ space, we can store the run-length compressed FM-index (RLFM)-index~\cite{makinen05rle}. 
Given $\SA[i]$, RLFM can recover $T[i-1]$ in $\Oh{\lg n}$ time. 
By storing RLFM-index in both directions, we can sequentially extract characters in \Oh{\lg n} time, 
which we use to match the next factor in \CDAWG{} --- remembering that each \ST{} node representation also stores the corresponding \SA{} range.

Let us recall that for computing a factor $F_x = T[\dst_x..\dst_x+|F_x|-1]$, the only thing left undone is to find its \Stab{} interval.
For that, we stipulate the invariant that when computing $F_x$, we have selected the \SA{} leaf $\lambda$ whose suffix number is $\dst_x$.
To ensure this invariant for $F_{x+1}$, we call \fnSuffixLink($\lambda,|F_x|$) to obtain the needed \SA{} leaf.
Finally, we find the locus of $F_x$ by \fnStrAncestor($\lambda, |F_x|$). 
Since each \ST{} node stores its \SA{} range, we have all the information to add the interval of $F_x$ to \Stab{}, and we are done.
The time complexity is dominated by the \ST{} simulation of \CDAWG{}.

\begin{theorem}
  For a text $T$ of length $n$, 
there exists a data structure of size \Oh{e + \bar{e}}, which can, given an interval $\intervalI \subseteq [1..n]$,
compute the LZ78 factorization of $T[\intervalI]$ in $\Oh{z \lg n}$ time with $\Oh{z}$ extra space, where $z$ is the number of computed factors.
\end{theorem}

\subsection{Second Approach: Climbing Up} \label{ssse:find_path}

In what follows, we show how to get rid of the dependency on the \ST{} simulation, 
which costs us $\Oh{\lg n}$ time per query and makes it necessary to also store the CDAWG of the inverted text.
Instead of simulating the \ST{} leaf with suffix number $\dst_x$ for computing factor $F_x$,
we select \varSink{} and search for a path to \varRoot{} of length $\ell := n - \dst_x + 1$.
This also means that instead of the top-down traversals as in the previous subsection, 
we climb up \CDAWG{} from \varSink{}.
To this end, we use the centroid path decomposition and some definitions.

\myblock{Centroid Path Decomposition}
By applying the centroid path decomposition on \ST{}, we obtain a centroid-path decomposed tree whose nodes are the heavy paths of \ST{} and its edges the remaining light \ST{} edges.
Each \varRoot{}-leaf path in the centroid-path decomposed tree has a length of \Oh{\lg n}.
\cite{belazzougui17cdawg} observed that the \CDAWG{} edges corresponding to the \ST{} heavy edges form a spanning tree of \CDAWG{}.
We apply the centroid path decomposition to the spanning tree of heavy edges again.
We denote the heavy edges obtained by the second centroid path decomposition as the heavy edges of the CDAWG, and all other edges of the CDAWG as the light edges.
After the second centroid path decomposition, the heavy edges form a set of disjoint paths, and each \varRoot{}-\varSink{} path in \CDAWG{} visits at most $2 \lg n \in \Oh{\lg n}$ light edges.
\cref{fig_st_lztrie_cdawg} gives an example of the centroid path decomposition and the correspondence between \ST{} and \CDAWG{}.

To speed up the \CDAWG{} traversal for the factorization computation,
we want to skip heavy edges. 
For that, we accumulate the information about LZ nodes of all heavy nodes in a heavy path $P$ and store this information directly in $P$ so that we only need to query a heavy path instead of all its heavy nodes.
A linear \varSink{}-\varRoot{} traversal in \CDAWG{} thus visits \Oh{\lg n} light nodes and heavy paths.
We can perform this traversal efficiently with some preprocessing:

\myblock{Node Lengths}
Let $\fnLen(u)$ for a \CDAWG{} node $u$ denote the set of the string lengths of all \varRoot{}-$u$ paths in \CDAWG{}. 
Actually, the set $\fnLen(u)$ is an interval. This can be seen as follows: if there are \varRoot{}-$u$ paths with labels $X$ and $Y$ for $X \in \Sigma^*$ and $Y$ being a suffix of $X$,
then any suffix $Z$ of $X$ longer than $Y$ has the same occurrences as $X$ and $Y$ in $T$, implying that these occurrences all follow the same characters, and therefore we can also reach $u$ from \varRoot{} by reading $Z$.
As a consequence, we can represent $\fnLen(u)$ in \Oh{1} words by using both interval ends,
and augment each \CDAWG{} node $u$ with $\fnLen(u)$ without violating our space budget.

\myblock{Node Distances}
For two \CDAWG{} nodes $u$ and $v$ on the same heavy path,
let $\fnDist(u,v)$ be their string depth distance, which is well-defined because either $u$ is the parent of $v$ or vice versa (otherwise they cannot belong to the same heavy path).

\myblock{Upward Navigation}
Recall that our aim is, after determining a factor $F_x = T[\dst_x..\dst_x+|F_x|-1]$ with \Stab{}, to find its interval for indexing $F_x$ with \Stab{}.
For that, we climb up \CDAWG{} from \varSink{} and search a \varRoot{}-\varSink{} path $P$ of length $\ell := n-\dst_x+1$, which is the string depth of the \ST{} leaf having suffix number $\dst_x$.
Such a path $P$ is uniquely defined since the \ST{} nodes collapsed to a \CDAWG{} node have all distinct string depths.
In particular, \ST{} nodes with the same string depth cannot have isomorphic subtrees, and therefore no two \varRoot{}-$v$ paths can share the same length (substituting $v$ with non-root \ST{} nodes).

For upward navigation,
we augment each node $v$ with a binary search tree $B_v$.
For each parent $u$ of $v$ connected by a light edge $(u,v)$, we store $(u,v)$ with key $\min(\fnLen(u)) + c(u,v)$ in $B_v$,
where $c(u,v)$ is the number of characters on the edge $(u,v)$.
With $B_v$, we can find the last edge $(u,v)$ of the \varRoot{}-$v$ path $P$ of string length~$\ell$ in $\Oh{\lg e}$ time.
After climbing up to~$u$,
the remaining prefix of $P$ is a \varRoot{}-$u$ path $P'$ of string length $\ell-c(u,v)$.

Now, a \CDAWG{} ancestor $u$ of $v$ in the same heavy path can be a node in $P$ if and only if $\ell - \fnDist(u,v) \in \fnLen(u)$.
Finding the highest possible such ancestor can be done with an exponential search in \Oh{\lg e} time.
We end up with a \CDAWG{} ancestor $u$ of $v$ in $P$ that is connected to its parent node $w$ in $P$ via a light edge (or $u = \varRoot$, and we terminate the traversal).
We can find $w$ with $B_u$, and recurse on $w$ belonging to another heavy path closer to the root node.
In total, we visit \Oh{\min(\lg n, e)} = \Oh{\lg n} heavy paths and light nodes.
On each heavy path or light node that we process, we spend \Oh{\lg e} time.
Thus, the total time per factor is \Oh{\lg n \lg e}.

\myblock{Finding the \SA{} Range}
Given we process factor $F_x$, we use the above procedure to find the \ST{} locus of $F_x$ represented by \CDAWG{}.
For that, we stop climbing when we reach the shortest path $P$ with a string length of at least $|F_x|$.
However, unlike the previous approach, we do not have the \SA{} ranges at hand. 
To compute them, we perform the following pre-computation step:
We let each \CDAWG{} node store (a) the number of \ST{} leaves in the subtree rooted at one of its collapsed \ST{} nodes (this is well defined because all these collapsed \ST{} nodes have the same tree topology)
and (b) the number of \ST{} leaves of its lexicographically preceding sibling nodes, which we call the \teigi{aggregated \CDAWG{} value}.
Additionally, each heavy path stores from bottom up the prefix-sums of the aggregated \CDAWG{} values of the nodes such that we can get for the $i$-th node on a heavy path the number of all leaves of all lexicographically preceding siblings of the descendant nodes of the $i$-th node belonging to the same heavy path.
This whole pre-preprocessing helps us find the \SA{} range of $F_x$ as follows:
We use a counter $c$ that accumulates the leftmost border of the \SA{} range we want to compute. 
For that, we increment $c$ when climbing up to a light node by its aggregated \CDAWG{} value.
Additionally, when we leave a heavy node, we use the prefix-sum stored in its respective heavy path to perform the computation in constant time per light node or heavy path.
When we reach the \CDAWG{} node $v$ representing the locus of $F_x$, $c$ gives us the left border of the \SA{} range we want to compute.
However, the length of this \SA{} range is given by the subtree size stored in $v$.
This concludes our algorithm.

\myblock{Speeding Up by Interval-Biased Search Trees}
The above time can be improved from \Oh{\lg n \lg e} to \Oh{\lg n}
by implementing (a) $B_v$ and (b) the exponential search in each heavy path with \teigi{interval-biased search trees}.

\begin{lemma}[{\cite[Lemma~3.1]{bille15randomaccess}}]\label{lemIntervalBiasedSearch}
Given a sequence of integers $\ell_1 \le \cdots \le \ell_m$ from a universe $[0..u]$, 
the interval-biased search tree is a data structure of $\Oh{m}$ space that can compute, 
for an integer $p$ given a query time,
the predecessor of $p$ in $\Oh{\lg (u/x)}$ time,
where $x = \fnSuccessor(p) - \fnPredecessor(p)$ is the difference between the predecessor $\fnPredecessor(p)$ and successor $\fnSuccessor(p)$ of $p$ in $\{\ell_1,\ldots,\ell_m\}$.
\end{lemma}
We note that there are faster predecessor data structures with time related to the distance of the query element to the predecessor such as~\cite{ehrhardt17delta,belazzougui12predecessor},
which however do not improve the total running time, which is dominated by the number of nodes 
$\Oh{\lg n}$ we visit.

For the former (a), denoting $B_v$ as $B_\cdot$ for any node $v$,
during a \varSink{}-\varRoot{} traversal, a query of $B_\cdot$ always leads us to a higher node $v$ such that the next search in $B_\cdot$ is bounded by $\max(\fnLen(v))$, 
and therefore the query times in \cref{lemIntervalBiasedSearch} lead to a telescoping sum of $\Oh{\lg n}$ total time.

For the latter (i.e, (b) the heavy paths), we let each heavy path maintain an interval-biased search tree storing its \CDAWG{} nodes.
A node $u$ is stored with the key $\fnDist(u,v')$, where $v'$ is the deepest node in the heavy path.
At query time, we have the desired path-length $\ell$ and $\fnLen(u) = [\min(\fnLen(u))..\max(\fnLen(u))]$ available such that we can query for the highest node $u_1$
with $\fnDist(u_1,v) = \fnDist(u_1,v') - \fnDist(v,v') \le \ell - \min(\fnLen(u_1))$,
i.e., $\fnDist(u_1,v') \le \ell - \min(\fnLen(u_1)) + \fnDist(v,v')$
and the highest node $u_2$ with $\fnDist(u_2,v') \ge \ell - \max(\fnLen(u_2)) + \fnDist(v,v')$.
Then the deepest node among $u_1$ and $u_2$ is the highest ancestor of $v$ that is still in $P$ and is a member of the same heavy edge.
The time complexity forms like for (a) a similar telescoping sum if we add to each key $\fnDist(u,v')$ the maximum depth of a heavy path such that each heavy path visit shrinks the search domain to be upper bounded by the last obtained key.

\begin{theorem}\label{thmFinalResult}
  For a text $T$ of length $n$, 
there exists a data structure of size \Oh{e}, which can, given an interval $\intervalI \subseteq [1..n]$,
compute the LZ78 factorization of $T[\intervalI]$ in $\Oh{z \lg n + z \lg z} \subseteq \Oh{z \lg n}$ time and $\Oh{z}$ extra space, where $z$ is the number of computed factors.
\end{theorem}

\subsection{Replacing CDAWG by Other Data Structures}
Not only the CDAWG-based index mentioned above, but other compressed indexes can also be used for substring compression.  
To make this possible, we revisit the algorithm described in \Cref{FirstAlg} and identify the core operations it requires from the underlying data structure.  
To compute a new factor $F_x = T[\dst_x..\dst_x+|F_x|-1]$, 
we first find the \ST{} leaf $\lambda$ that represents $T[\dst_x..]$.
Note that $\lambda$ is the lexicographically $\ISA[\dst_x]$-th smallest suffix of $T$.
Second, we compute the lowest LZ78 node $u$ on the root-$\lambda$ path.
We find $u$ by actually performing a stabbing-max query on \Stab{} with value $\ISA[\dst_x]$.
Finally, we compute the \SA{} range $R$ corresponding to the locus of $F_x$ and insert $R$ into \Stab{}.
We observe that the \SA{} range $R$ computed by the above procedure is the same as the \SA{} range corresponding to $T[\dst_x..\dst_x + |F_x| - 1]$.
In general, our algorithm requires the following operations:
\begin{itemize}
    \item[(a)] compute $\ISA[i]$,
    \item[(b)] access $T[i]$,
    \item[(c)] compute the \SA{} range corresponds to a substring $T[x..y]$ for $1 \le x \le y \le n$, and
    \item[(d)] perform a stabbing-max query or update on a dynamic set of intervals.
\end{itemize}
Therefore, any data structure supporting (a)-(c) can take over the role of suffix trees or \CDAWG{}s for substring compression with our algorithm.

Here, we show that an RLBWT-based compressed index satisfies the above properties.
There is a compressed data structure that supports typical suffix tree operations in $\Oh{r \lg \frac{n}{r}}$ space~\cite{gagie20fully}.
It supports
\begin{itemize}
	\item random access on the text and on the inverse suffix array in $\Oh{\lg \frac{n}{r}}$ time, and 
	\item computing $\fnPrevSmaller(x, d) = \max \left( \{ 0 \} \cup \{ 1 \leq y < x \mid \LCP[y] < d \} \right)$ and \\$\fnNextSmaller(x, d) = \min \left( \{ n \} \cup \{ x \leq y < n \mid \LCP[y] < d\}\right)$ in $\Oh{\frac{\lg n}{\lg \lg n} + \lg \frac{n}{r}}$ time.
\end{itemize}
It thus supports operations (a) and (b).
We can support (c) by computing $[\fnPrevSmaller(\ISA[x], y - x + 1), \fnNextSmaller(\ISA[x], y - x + 1) - 1]$.
Therefore, by replacing \CDAWG{} with this data structure, we can obtain the following result.
\begin{theorem} \label{thmRLBWT}
For a text $T$ of length $n$,
there exists a data structure of size $\Oh{r \lg \frac{n}{r}}$, which can, given an interval $\intervalI \subseteq [1..n]$,
compute the LZ78 factorization of $T[\intervalI]$ in $\Oh{z \left(\lg \lg \frac{r}{\lg n} + \lg \frac{n}{r} + \lg z \right)}$ time with $\Oh{z}$ extra space, where $z$ is the number of computed factors and $r$ is the number of BWT-runs of $T$.
\end{theorem}

Furthermore, we also show 
that the \teigi{$\delta$-index}~\cite{kempa23collapsing} also satisfies the above requirements.
The $\delta$-index can be stored in $\Oh{\delta \lg \frac{n \lg \sigma}{\delta \lg n}}$ space, and supports random access for the text and $\ISA$ in $\Oh{\log^{4+\varepsilon} n}$ time, where $\varepsilon > 0$ is a given constant.
Furthermore, it also allows computing the length of the longest common prefix between any two suffixes of the text in $\Oh{\lg n}$ time.
Since it supports LCE operations, we can simulate $\fnPrevSmaller(x, d)$ and $\fnNextSmaller(x, d)$ 
using the $\delta$-index by combining binary search and LCE queries. 
Therefore, we can use it instead of \CDAWG{} and the RLBWT-based compressed index for the LZ78 substring compression problem.
With the $\delta$-index, we obtain the following result.

\begin{theorem} \label{thmDelta}
For a text $T$ of length $n$, there exists a data structure of size $\Oh{\delta \lg \frac{n \lg \sigma}{\delta \lg n}}$, which, given an interval $\intervalI \subseteq [1..n]$, can
compute the LZ78 factorization of $T[\intervalI]$ in $\Oh{z \lg^{4+\varepsilon} n}$ time with $\Oh{z}$ extra space, where $z$ is the number of factors computed, and $\delta$ is the substring complexity of $T$, and $\varepsilon > 0$ is a constant.
\end{theorem}
Note that we focus solely on the theoretical result because there is no practical implementation of this data structure, and it is not designed with practical performance in mind.

\subsection{Extension for LZMW and LZD}
\begin{figure}[t]
\centering
\includegraphics[width=0.8\textwidth]{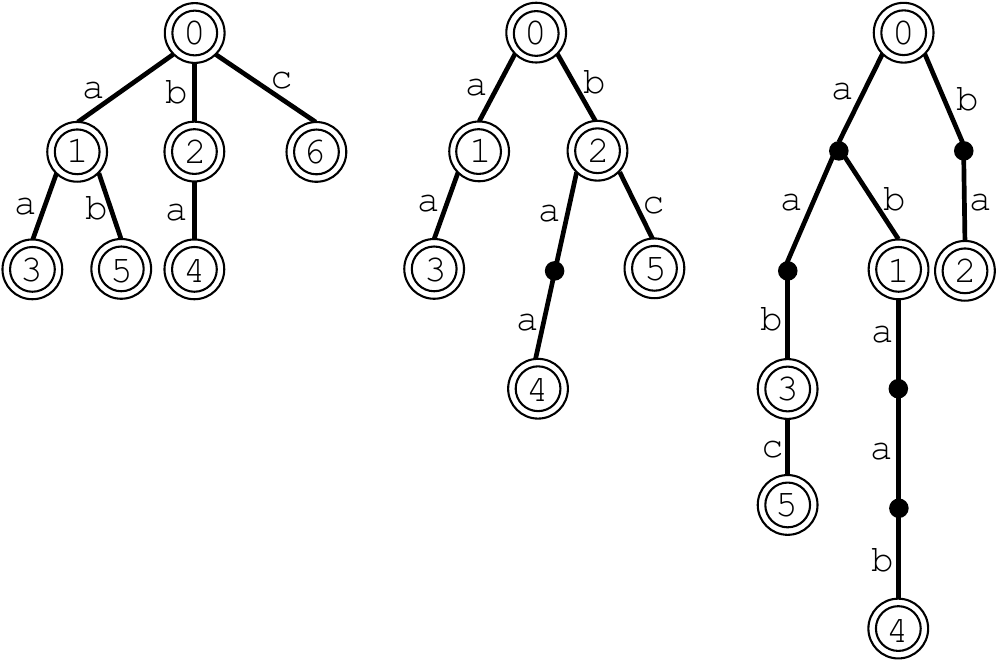} 
\caption{%
    The LZ78 (left), LZD (center), and LZMW (right) tries for $T = \texttt{abaabaabc}$, the same string as in~\cref{fig_factorizations}.  
    Vertices shown as double circles represent nodes corresponding to factors in each LZ trie, while black dots mark intermediate nodes that do not correspond to any factor.
    The integer inside each node indicates the index of the corresponding factor.  
    In the LZ78 and LZD tries, a node labeled $i$ represents the factor $F_i$.  
    In the LZMW trie, a node labeled $i$ represents the concatenation of two consecutive factors, namely the string $F_i F_{i+1}$, except for the special case when $i = 0$.
}
\label{fig_lztries}
\end{figure}

In this section, we show the compressed-space substring compression algorithms for LZMW and LZD factorization.
The main difference between these factorizations and LZ78 lies in the structure of their corresponding LZ tries.  
\cref{fig_lztries} illustrates the LZ tries considered in this work.  
While the standard LZ78 trie contains only nodes that correspond to factors, the LZD and LZMW tries also include intermediate nodes that do not represent any factor.
As a result, the total number of nodes in these tries may exceed the number of factors.  
Therefore, maintaining the LZ tries in a straightforward manner does not guarantee good space complexity.  
However, our approach avoids this issue by not maintaining the trie topology explicitly.  
Instead, we associate each factor with an interval in the suffix tree, which allows us to sidestep the overhead caused by the structural differences in the LZ tries.

\myblock{LZD Factorization}  
We first review the original substring compression algorithm for LZD proposed in~\cite{koppl24lz78derivates}.  
Suppose we want to compute a new factor $F_i = G_1 G_2$.  
In the algorithm, the first half $G_1$ is computed using the LZ trie consisting of $\{ F_0, \dots, F_{i-1} \}$, which is superimposed on the suffix tree.  
The second half $G_2$ is then computed in the same manner as the first.  
Finally, a new node corresponding to the new factor $F_i$ is inserted into the LZ trie.

Since the procedures for computing both $G_1$ and $G_2$ are essentially the same as in the LZ78 algorithm,  
we can directly apply the compressed-space algorithm developed for LZ78.  
In addition, the method for computing the new \SA{} range described in~\cref{ssse:find_path} can also be reused.  
Therefore, by using compressed-space data structures for LZ78 substring compression, we obtain the same performance guarantees for LZD.

\myblock{LZMW Factorization}  
We next consider the LZMW substring compression problem.  
Unlike the LZ78 and LZD tries, the LZMW trie consists of all concatenations of two consecutive factors.  
In the LZMW substring compression algorithm, we superimpose the LZMW trie onto the suffix tree, in the same way as done for LZ78.  
Using this superimposed structure, we find the longest matching factor starting at position $p$ in $T$ by identifying the deepest mark corresponding to an LZMW factor along the path representing $T[p..n]$.  
Since this procedure is nearly identical to that of the LZ78 substring compression, our compressed-space method can be naturally extended to support LZMW factorization as well.  
Thus, we can compute the LZMW substring compression using the same data structures.

Therefore, using the methods of \cref{thmFinalResult}, \cref{thmRLBWT}, and \cref{thmDelta}, we can obtain these results.

\begin{theorem}\label{thmVariants}
  For a text $T$ of length $n$ over an alphabet of size $\sigma$, there exist data structures that, given an interval $\intervalI \subseteq [1..n]$, can compute the LZMW or LZD factorization of $T[\intervalI]$ under the following space and time complexities: \begin{itemize}
    \item A structure of size $\Oh{e}$ takes $\Oh{z \lg n + z \lg z} = \Oh{z \lg n}$ time with $\Oh{z}$ extra space.
    \item A structure of size $\Oh{r \lg \frac{n}{r}}$ takes $\Oh{z \left(\lg \lg \frac{r}{\lg n} + \lg \frac{n}{r} + \lg z\right)} \subseteq \Oh{z \lg n}$ time with $\Oh{z}$ extra space.
    \item A structure of size $\Oh{\delta \lg \frac{n \lg \sigma}{\delta \lg n}}$ takes $\Oh{z \lg^{4+\varepsilon} n}$ time with $\Oh{z}$ extra space.
  \end{itemize}
  Here, $z$ is the number of computed factors, and $r$ is the number of BWT-runs of $T$, $\delta$ is the substring complexity of $T$, and $\varepsilon > 0$ is a constant.
\end{theorem}

\section{Experiments}
In what follows, we empirically evaluate CDAWG-based and RLBWT-based indexes on real-world text strings for computing LZ78 substring compression.
To do this, we first highlight the details of our implementation in \cref{exp_impl}.
Subsequently, we describe our experimental settings in \cref{exp_settings}.
Finally, we report the memory consumption, running time, and the distribution of the number of edges on paths between the \CDAWG{} root and its sink in \cref{exp_results}.

\subsection{Deviation from Theory} \label{exp_impl}
We implement a simplification of our CDAWG-based index proposed in \cref{ssse:find_path}.
In particular, we omit the centroid path decomposition because we empirically observed that the average number of edges on the \varRoot{}-\varSink{} paths is small in our datasets.
We will discuss this observed phenomenon in detail in \cref{exp_results}.
To reduce complexity, we implemented the branches of each internal node, instead of an interval-biased search tree, by a sorted list on which we do a binary search to find the edge with the right label.
We also deviate from theory in the implementation of the stabbing-max data structure, for which we use splay trees~\cite{SleatorSelfAdjusting}.
With a splay tree built on $z$ intervals, the times to answer a query or add an interval are $\Oh{\lg z}$ amortized each, and the space is $\Oh{z}$ words.
Therefore, replacing the original data structure with splay trees does not worsen the space of our index and keeps the time within $\Oh{z \lg z}$.
Splay trees provide fast access to frequent elements by rearranging their structure adaptively on each query, and thus can exploit skewed distributions unlike common balanced trees such as AVL trees.
The reason for using splay trees is that vertices of the splay tree are sequentially inserted at positions adjacent to the vertex that becomes the root of the splay tree by the previous query.
By doing so, chances are high that a splay tree query only involves the very upper part of the tree, making the implementation practically fast in most cases.

For comparison, we also implement a simplification of the \ST{}-based index.
Our implementation differs from the method proposed in \cite{koppl21nonoverlapping} in the following two points:
(i) we omit the weighted-ancestor data structure, and, (ii) we use a stabbing-max data structure instead of a lowest marked ancestor data structure.
The first change is because the average number of edges on the \varRoot{}-leaf paths of \ST{} is small on our datasets, similar to the \varRoot{}-\varSink{} paths of \CDAWG{}.
The second change aims to reduce memory consumption.
The stabbing-max data structure requires only $\Oh{z}$ space, whereas the lowest marked ancestor data structure requires $\Oh{n}$ space in addition to \ST{}.

We also implement a \RLBWT{}-based index.
This implementation is almost the same as the method proposed in \cite{gagie18bwt}, except for the implementation of the predecessor structure used for computing $\fnPrevSmaller(x, d)$ and $\fnNextSmaller(x, d)$.
In the original method, they split an LCP array of length $n$ into $\Oh{r}$ blocks and maintain each block in grammar-compressed form.
Finally, they use a predecessor structure to determine the block containing the position $x$.
Since each block has a tree structure of height $\Oh{\lg \frac{n}{r}}$ and the used predecessor structure needs $\Oh{\lg \lg \frac{r}{\lg n}}$ query time, 
the time complexity for computing $\fnPrevSmaller$ and $\fnNextSmaller$ is $\Oh{\lg \lg \frac{r}{\lg n} + \lg \frac{n}{r}}$.
However, the predecessor structure mentioned by the authors focus only on the theoretical result and is not going to be implemented.
For simplification, in our implementation, we concatenate all blocks and construct a grammar tree of height $\Oh{\lg n}$.
Therefore, our solution answers these queries in $\Oh{z \lg n}$ time.

\myblock{Implementation Details}
\newcommand{\NodeArray}{\ensuremath{A_{\mathup{V}}}}
\newcommand{\EdgeArray}{\ensuremath{A_{\mathup{E}}}}
We maintain the nodes and the edges of \CDAWG{} separately in two arrays~\NodeArray{} and~\EdgeArray{}.
We store nodes in \NodeArray{} in an arbitrary order,
while we store edges in \EdgeArray{} in a sorted order based on two criteria.
First, we partially sort the edges in groups sorted by the \NodeArray{} index of the connecting child node.
Second, for a fixed child node $v$, an edge $(u,v)$ is sorted by the key $\min(\fnLen(u)) + c(u,v)$ within its groups of edges sharing the same child node $v$.
This arrangement makes it possible to perform binary search on the edge array for simulating the binary search trees $B_v$, which we here no longer need.
Given a node $v$, to jump into the range $[\ell..r]$ of the edge array of edges connecting to $v$ for querying $B_v$, we let $v$ store $\ell$.
We can do so by letting node $\NodeArray[i]$
store 
(V1) the sum of the number of children over all preceding nodes (summing up the number of children of node $\NodeArray[j]$ for each $j \in [1..i-1]$).
We then also know the right end of the interval $[\ell..r]$ by querying the subsequent node in the \NodeArray{}.
Additionally, each node stores 
(V2) $\max(\fnLen(v))$ and 
(V3) the number of paths from $v$ to the sink.
An edge $(u,v)$ from a node~$u$ to its child~$v$ is represented as a tuple of three integers:
(E1) the index of $u$'s entry in \NodeArray{}, 
(E2) the string length of $(u,v)$, and 
(E3) the prefix sum of $u$'s aggregated \CDAWG{} values (defined in \cref{ssse:find_path}).
Therefore, both a node and an edge store three integers each.
Following \cref{ssse:find_path}, we use these integers as follows:
(E1) to select the parent node of $v$ returned by $B_v$,
(E2)  to simulate a query on $B_v$ via binary search with (V2), 
and to compute the string depth of the updated path when moving upward to the returned parent, and
(E3) with (V3) to determine the \SA{} range of the factor we want to compute.
In addition, we store the first character of each edge label for each edge incident to \varRoot{} to provide efficient random access to $T$.
To restore $T[i]$ from \CDAWG{}, we first compute the path $(e_1, e_2, \dots, e_k)$ representing $T[i..n]$.
Then, we obtain $T[i]$ by taking the first character of the edge label of $e_1$.
Storing these labels takes $\sigma \lg \sigma$ bits in total.
Therefore, the overall memory consumption is $3 p (|V| + |E|) + \sigma \lg \sigma$ bits, 
where $p \in \Om{\lg n}$ is the size of an integer in bits.

Our suffix tree consists of three parts:
an array of nodes, an array of pointers to all leaves sorted by their suffix numbers, 
and the raw input text.
Each node $v$ stores its string depth,
the index of $v$'s parent node in the node array, 
and $v$'s \SA{} range.
With the node array and the pointers to the leaves, 
we can determine the \SA{} range for \Stab{}.
We do not need $\SA$ because for computing the LZ78 substring compression,
we only need to compute the \SA{} range corresponding to an LZ78 factor, not the actual \SA{} values.
The total memory consumption is $4p|V| + n \lg \sigma$ bits.

Our index based on \RLBWT{} consists of three data structures: a block tree to access $T[i]$, a block tree for accessing $\ISA[i]$, and a grammar-based index to compute $\fnPrevSmaller(x, d)$ and $\fnNextSmaller(x, d)$.
We omit the details of the implementation, but the overall data structure is almost the same as the structure of \cite{gagie18bwt}, except for the predecessor structure.
Each of the tree data structures consumes $\Oh{rp \lg \frac{n}{r}}$ bits of space, where $p \in \Om{\lg n}$ is the size of an integer in bits.

\begin{table}[t]
    \centering
    \caption{%
    Sizes and memory usage of \CDAWG{}, \ST{} and \RLBWT{} of each dataset.
		Memory is measured in mebibytes (MiB).
    \ST{} has approximately $2n$ vertices and edges regardless of the dataset. 
    }
    \label{table_text_ratio}
    \begin{tabular}{l rr rr rrr}
        \toprule
        &  
        \multicolumn{2}{c}{\CDAWG{} size}
        &
        \multicolumn{2}{c}{\RLBWT{} size}
        &
        \multicolumn{3}{c}{memory usage}
        \\
        \cmidrule(lr){2-3}
        \cmidrule(lr){4-5}
        \cmidrule(lr){6-8}
        dataset
        &  
        $e$ &
        $e / n$ &
        $r \lg \frac{n}{r}$ &
        $r \lg \frac{n}{r} / n$ &
        \ST{} & \CDAWG{} & \RLBWT{}
        \\ \midrule
        \textsc{sources}        & 66.33e6 & 0.494 
        & 65.74e6 & 0.489
        & 3970.6 & 984.5 & 1764.9 \\
        \textsc{dna}     & 178.91e6 & 1.333
        & 56.74e6 & 0.422
        & 4069.3 & 2830.2 & 1258.1 \\
        \textsc{english}    & 102.21e6 & 0.761 
        & 71.21e6 & 0.531
        & 3920.0 & 1518.6 & 1809.5 \\
        \textsc{fib}    & 74 & 5.513e-7 
        & 453.6 & 3.38e-6
        & 4736.0 & 1.28e-3 & 2.76e-2 \\
        \bottomrule
    \end{tabular}
\end{table}

\begin{figure}[t]
\makebox[\textwidth][c]{
    \includegraphics[width=\textwidth]{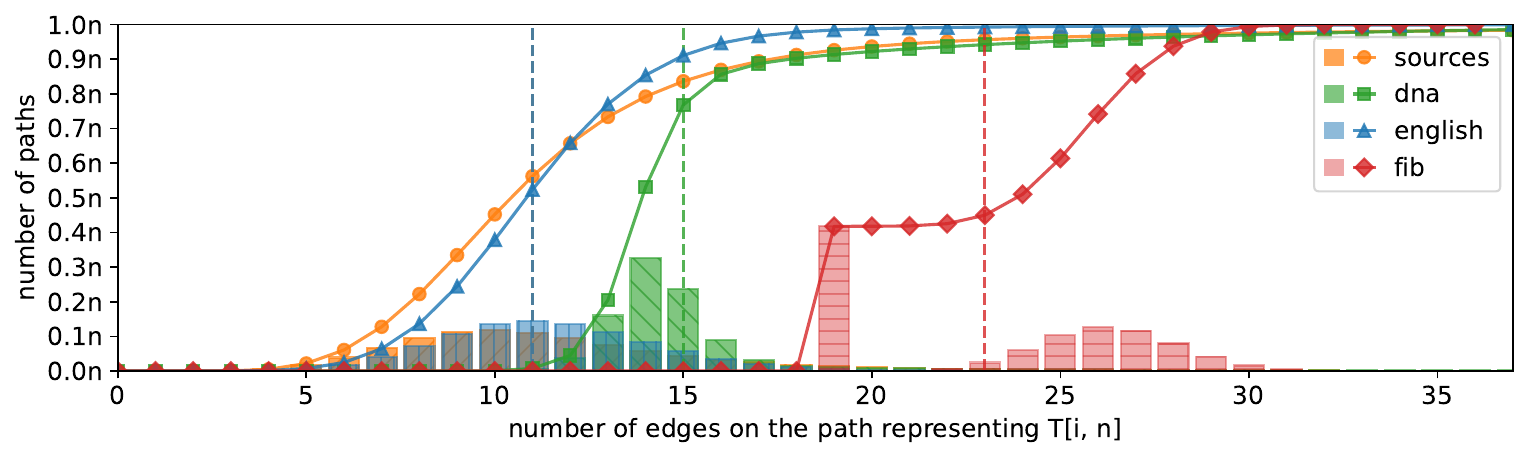} 
}
\caption{%
	The histogram of the number of edges on all paths between \varRoot{} and \varSink{}.
The dashed vertical line for each dataset represents the average number of edges on all \varRoot{}-\varSink{} paths.
The curve represents the cumulative sum of the histogram.
Note that the number of all \varRoot{}-\varSink{} paths is $n=134,217,728$ for all datasets.
}
\label{fig_number_of_edge_distribution}
\end{figure}
\begin{table}[t]
    \centering
    \caption{%
    The alphabet size and
    repetitiveness measures on the first 128MiB of each dataset ($n = 134,217,728$).
    $\sigma, e, r, z_{77}$ and $z_{78}$ represent the alphabet size, the number of edges in the CDAWG, the number of runs of Burrows--Wheeler transform, and the number of factors of LZ77 and LZ78 factorization, respectively.
	Note that $e \in \Omm{\max\{ r, z_{77} \}}$ holds for any text~\cite{navarro21indexing1}.}
    \label{table_text_repetitive_measure}
    \setlength{\tabcolsep}{1em}
    \begin{tabular}{l r rr rr}
        \toprule
        dataset &  
        $\sigma$ &
        $e$ &
        $r$ &
        $z_{77}$ &
        $z_{78}$
        \\ \midrule
        \textsc{sources} & 227 & 663278 & 31303555 & 7816156 & 13811755 \\
        \textsc{dna} & 16 & 178908741 & 84071820 & 9284690 & 10825116 \\
        \textsc{english} & 218 & 102211137 & 48367053 & 9639620 & 14219552 \\
        \textsc{fib} & 2 & 74 & 20 & 41 & 267813 \\
        \bottomrule
    \end{tabular}
\end{table}

\subsection{Experimental Settings} \label{exp_settings}
We have implemented our LZ78 substring compression algorithms in C++.
The source code is available at \url{https://github.com/shibh308/CDAWG-LZ78}.
For simplification, we assume that the input is interpreted in the byte alphabet ($\lg \sigma$ = 8) and $n \le 2^{32}$ (thus $p = 32$).
\Cref{table_text_repetitive_measure} gives characteristics of the input texts used.

In one experiment instance, 
we construct the CDAWG, the ST, or the RLBWT-based index of an input text and answer some LZ78 substring compression queries.
As input texts, we used \textsc{sources}, \textsc{dna}, and \textsc{english} from the Pizza\&Chilli Corpus~\cite{ferragina08compressed},
and the length-$n$ prefix of the (infinite) Fibonacci string \textsc{fib}.
We note that the construction time of these indexes is $\Oh{n}$ in theory. However, since our implementation is not optimized for construction speed, we omit experiments evaluating construction time.
It is known that the CDAWG of the length-$n$ prefix of the Fibonacci string has only $\Oh{\lg n}$ edges~\cite{rytter06fibonacci}, and $r \in \Oh{\lg n}$ also holds from $r \in \Oh{e}$.
We fixed $n = 2^{27} = 134,217,728$, and generated our input texts by extracting
the first $2^{27}$ bytes (=128MiB) from each dataset from the text collection.

We compiled our source code with GCC 12.2.0 using the -O3 option,
and ran all experiments on a machine with Debian 12, Intel(R) Xeon(R) Platinum 8481C processor, and 64GiB of memory.

\begin{figure}[t]
    \centering
    \begin{minipage}[c]{0.6\textwidth}
        \includegraphics[width=\textwidth]{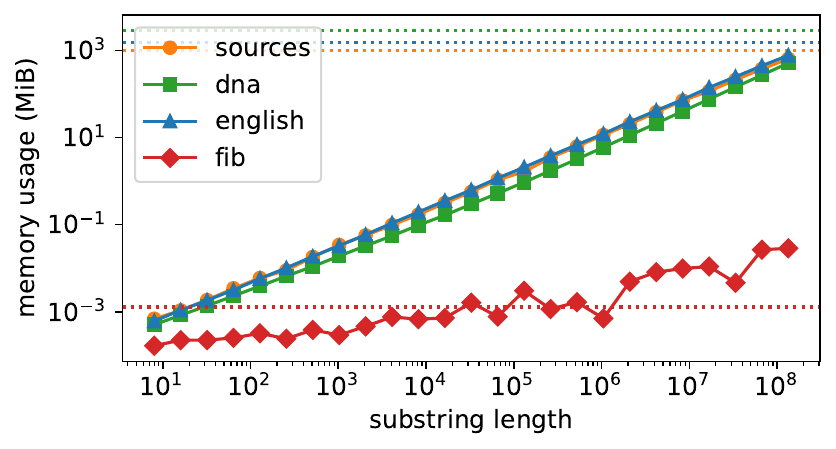} 
    \end{minipage}
    \hfill
    \begin{minipage}[c]{0.35\textwidth}
        \caption{Average memory usages of the stabbing-max data structure depicted by solid lines.
        The dotted line with the same respective color represents the memory usage of the CDAWG of the respective dataset.
        }
    \label{fig_stabbing_max_memory_usage}
    \end{minipage}
\end{figure}

\begin{figure}[ht]
    \centering
    \includegraphics[width=1.0\textwidth]{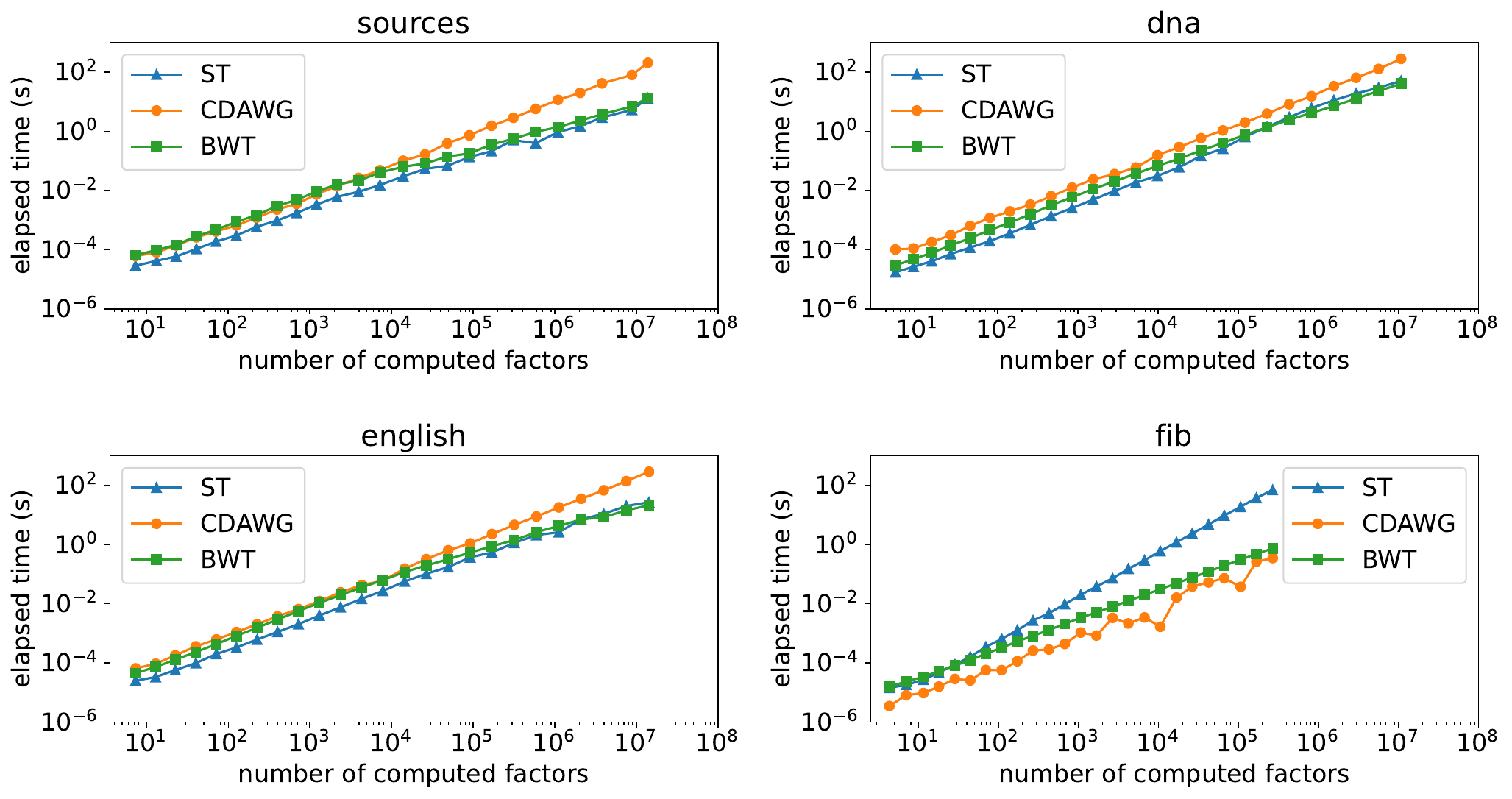}
    \caption{Average elapsed time for LZ78 substring compression with \ST{}, \CDAWG{}, and \RLBWT{}.}
    \label{fig_elapsed_time}

\centering
\includegraphics[width=1.0\textwidth]{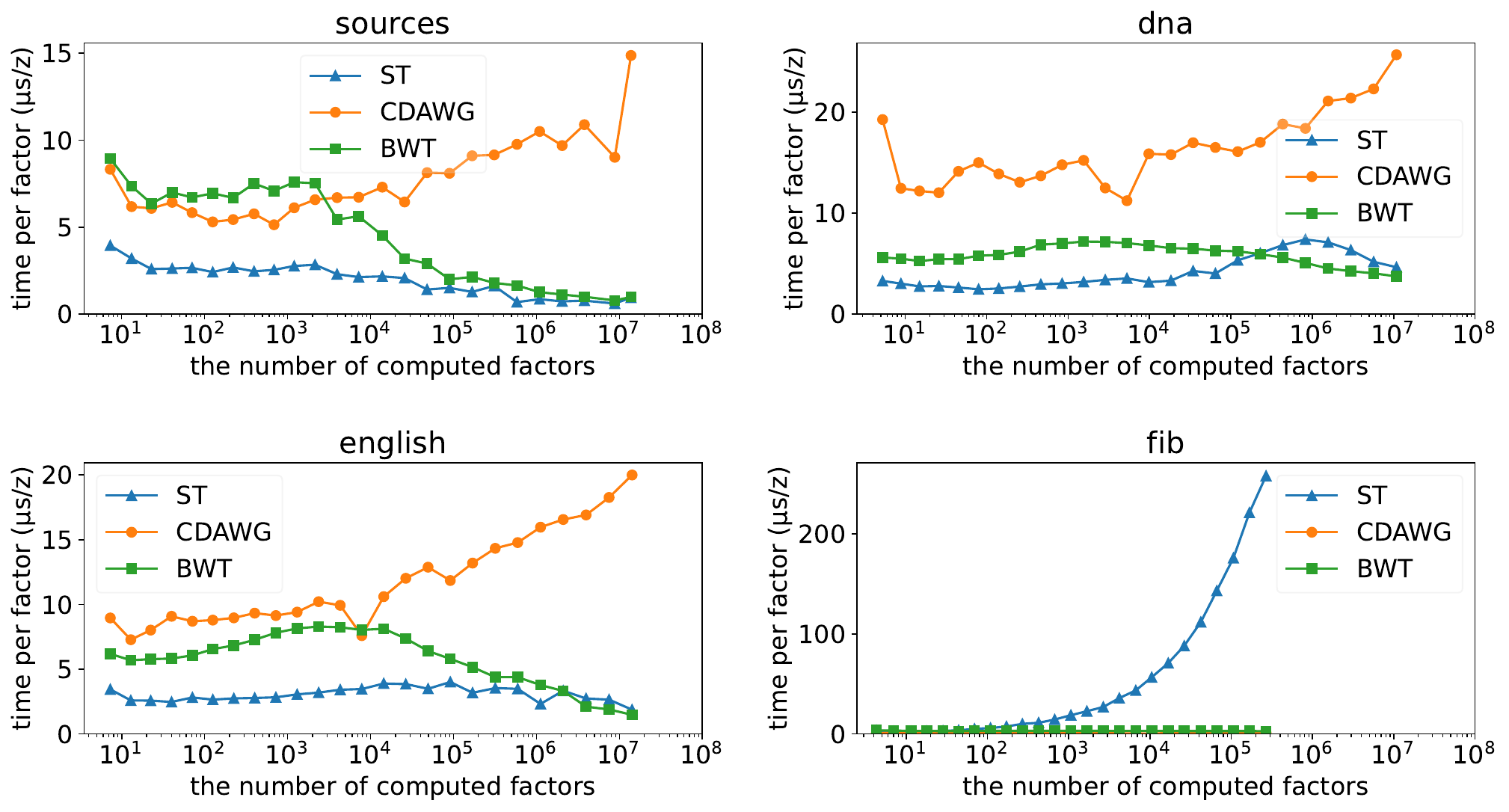} 
\caption{Average elapsed time divided by the number of computed factors for LZ78 substring compression with \ST{}, \CDAWG{}, and \RLBWT{}.}
\label{fig_elapsed_time_by_lz78_factor}
\end{figure}

We first construct the indexes for each text string and compute its memory consumption.
We also measure the distribution of the number of edges on the \varRoot{}-\varSink{} paths of \CDAWG{}.
Note that we did not measure the construction time because we did not focus on efficient construction.
After construction, 
we let them answer LZ78 substring compression queries.
For each $\alpha \in \{2^3, 2^4, \dots, 2^{27}\}$,
we choose ten substrings of length $\alpha$ from the text uniformly at random and compute the LZ78 compression of these substrings.
We calculate the average memory consumption of the stabbing-max data structure and the elapsed time excluding the maximum and minimum values.

\subsection{Experimental Results} \label{exp_results}

\cref{table_text_ratio} indicates the size and memory consumption of each approach.
For all datasets, \CDAWG{} and \RLBWT{} consume less memory than \ST{}.
Except for \textsc{dna}, \CDAWG{} consumes less space than \RLBWT{}.
However, it consumes more than twice the space of \RLBWT{} in the case of \textsc{dna}.
Furthermore, \CDAWG{} takes more space than the input itself because each edge and vertex consists of multiple integers.
Even more severe, the number of \CDAWG{} edges alone is higher than $n$ for \textsc{dna}.
For \textsc{fib}, both \CDAWG{} and \RLBWT{} compress the input text exponentially.
Especially, the memory usage of \CDAWG{} is about $10^5$ and $3.7 \times 10^6$ times less than the raw text and \ST{}, respectively.
%

\cref{fig_number_of_edge_distribution} shows the distribution of the number of edges on all \varRoot{}-\varSink{} paths.
We observe that the average number of edges on a \varRoot{}-\varSink{} path is about 10--15, and almost all paths have at most 20 edges in real-world datasets.
Therefore, we can regard the number of paths as almost $O(\lg n)$.
For plain CDAWGs without centroid path decomposition, path extraction can take $\Oh{n}$ time at worst.
However, from this result, 
we empirically constitute that such cases are rare in practice, and both with and without centroid path decomposition, the running times are almost the same.
Note that the average number of edges on the \varRoot{}-\varSink{} paths of \CDAWG{} is the same as the number of \varRoot{}-leaf paths of \ST{}, so this fact also applies to \ST{}.
%

\Cref{fig_stabbing_max_memory_usage} plots the memory usage of the stabbing-max data structures used.
Memory consumption increases almost linearly with the increase in the length~$\alpha$ of the queried substring.
The memory usage of the stabbing-max data structure is about 50--80\% lower than the memory consumption of \CDAWG{} except for \textsc{fib} even if $n = \alpha$ (i.e., the substring to compress in the whole text).
Therefore, for real-world datasets, the main memory bottleneck is the CDAWG\@.
%

%

\Cref{fig_elapsed_time} shows the elapsed time for LZ78 substring compression using \ST{}, \CDAWG{}, and \RLBWT{}.
\Cref{fig_elapsed_time_by_lz78_factor} shows the elapsed time divided by the number of factors computed.
The elapsed time increases almost linearly with the increase in the length of the queried substring.
On real-world datasets, \CDAWG{} is the slowest and \ST{} is the fastest for most cases.
The difference in computation time between \CDAWG{} and \ST{} is about 2-20 times.
This is because the traversal of the ST is simpler than that of the others.
In contrast, \CDAWG{} and \RLBWT{} computes the LZ78 substring compression about 5--300 times faster than \ST{} in \textsc{fib}, and \CDAWG{} is the fastest in this case.
We speculate that effective \CDAWG{}-based compression has a positive impact on cache-friendly memory access.

\section{Conclusion}
We propose a technique that allows us to compute the substring compression of LZ78, LZD, and LZMW in compressed space.
For that we introduce an abstract data type, whose implementations give various trade-offs, cf.~\cref{tabSolutions}.
We conducted experiments on different types of data, and empirically evaluated that our method performs LZ78 substring compression efficiently with space improvements compared to methods based on \ST{}.
For that, we slightly deviated from the theory by omitting the centroid path decomposition and sophisticated data structures.

There are several directions for future work.  
Reducing the number of required operations for substring compression and exploring smaller indexes that support substring compression is an interesting line of research.  
For instance, grammar-based compressed indexes are theoretically smaller than CDAWG-based ones, and support not only random access but also pattern matching queries.  
However, they do not support suffix array queries, which are essential in our current method for LZ78 substring compression.  
Developing a new substring compression technique that avoids relying on suffix array queries could allow the use of grammar-based indexes, resulting in a more space-efficient solution.

Substring compression for dynamic texts is another promising direction.  
When the text is modified, the compressed index must be updated accordingly.  
A straightforward approach would be to use known compressed indexes that support edit operations~\cite{nishimoto24dynamic,nishimoto25dynamic}, but most of these incur high computational costs.  
However, in our case, we only require the support for a limited set of operations.  
Thus, by focusing on this restricted set, it may be possible to significantly speed up the update process.

\myblock{Acknowledgments}
This work was supported by the JSPS KAKENHI Grant Numbers JP23H04378 and JP25K21150.

\clearpage


\bibliography{literature,references}

\end{document}